\documentclass[11pt,a4paper]{article}
\usepackage[T1]{fontenc}
\usepackage[utf8]{inputenc}
\usepackage{mathptmx}
\usepackage{xcolor}
\usepackage[margin=1in]{geometry}
\usepackage{amsmath,amssymb}
\usepackage{graphicx}
\usepackage{booktabs,array,ragged2e}
\usepackage[hyphens]{url}
\usepackage[colorlinks=true,linkcolor=blue!60!black,urlcolor=blue!60!black,citecolor=blue!60!black]{hyperref}
\usepackage{microtype}
\usepackage{caption}
\newcolumntype{L}[1]{>{\RaggedRight\arraybackslash}p{#1}}
\title{Memory Scarcity, Open Models, and the Restructuring of the AI Industry, 2026--2030\\[6pt]{\large A quantitative scenario analysis of inference economics, training-cost divergence, and infrastructure solvency}}
\author{Satoshi Matsuoka\\RIKEN Center for Computational Science (R-CCS)\\\texttt{matsu@acm.org}}
\date{July 8, 2026}
\begin{document}\maketitle
\begin{abstract}
This report examines how four simultaneous forces restructure the AI industry over 2026--2030: the historic DRAM/HBM price surge; the arrival of frontier-capable open-weight models exemplified by GLM-5.2; rapid inference-efficiency gains exemplified by near-Shannon-limit KV-cache compression (TurboQuant) and lightweight local runtimes (DwarfStar 4, DGX Spark-class hardware); and the entry of ``former AI labs'' --- Meta and xAI --- into the compute-resale market on the strength of fleets acquired before the memory repricing.

The central quantitative findings are as follows. First, the cost advantage of incumbent sunk fleets over new entrants is structural and does not close within the horizon: measured in dollars per petabyte of memory bandwidth delivered (the correct unit for bandwidth-bound decode), the entrant gap runs 3.2$\times$ in 2026, narrows to 1.9$\times$ in 2027, and then re-widens to roughly 3$\times$ (if HBM prices normalize in 2028) or above 4$\times$ (if the shortage persists to 2030), because the depreciation conveyor continuously delivers newly-amortized fleets to incumbents faster than hardware prices normalize. The advantage rotates among incumbents; it never transfers to entrants. Second, training economics bifurcate into a luxury tier and a mass tier: the cost of a frontier-class run grows to \$18B--\$38B by 2030 while replicating previous-frontier capability through reinforcement learning and distillation on open bases falls toward \$5M, a divergence from roughly 40$\times$ today to three-to-four orders of magnitude. Third, solvency of new infrastructure is confined to a narrow corridor: given the announced buildout and 30\%/yr bytes-per-token efficiency gains, aggregate token demand must sustain approximately 2$\times$ annual growth for four consecutive years, and premium (closed-model) pricing must remain sticky in absolute terms. Exiting the corridor on either axis concentrates impairments on peak-vintage capacity --- with the 2028--2029 delivery window, whose final investment decisions are being signed now, as the single most exposed tranche. Fourth, the strongest counterargument to the breakdown thesis is efficiency deceleration: KV compression is already near its information-theoretic limit, and if efficiency gains slow from 30\%/yr to 15\%/yr the solvency threshold falls from 1.9$\times$ to about 1.6$\times$ annual token growth, materially widening the corridor. Fifth, a greenfield entrant shipping custom silicon into a new datacenter removes the merchant margin but not the memory premium: the central outcome distribution is 25\% success, 34\% mediocrity, and 41\% loss, with the loss probability on deployed capital reducible to roughly 24\% through a staged go/no-go procedure whose gates are synchronized to the paper's standing tracker. Sovereign and institutional capacity is treated throughout as a third balance-sheet class outside the solvency corridor, and AI for Science and Industrial Innovations (AI4SIS) demand --- mission-funded and rationing-proof --- is identified as the corridor's inelastic floor.

Five scenarios organize the outcome space. The revised probabilities are: Rotating Landlord Oligopoly 25\%; Jevons Absorption 20\%; System-Layer Re-differentiation 18\%; Commoditization Crash 25\%; Geopolitical Bifurcation 12\%. The Crash case is no longer a tail: it is co-modal with the landlord outcome after accounting for demand-quality measurement error (9.1), token minimization and on-premises reallocation (9.2), the projection-vintage problem (9.3), and circular-finance fragility (10.1). However, the post-Q2 2026 regime break is not yet confirmed by a sufficient time series: pre-break projections are treated as optimistic bounds until two quarters of post-break data establish the new slope, and bandwidth demand peaking around 2028 is elevated from tail risk to co-equal stress case rather than sole base case. The paper's central conclusion is that the AI infrastructure boom is no longer underwritten by gross token growth alone: solvency now depends on monetized bandwidth demand, premium-price stickiness, and the ownership of vintages of memory-heavy capacity. The single most decision-relevant observable is realized token-demand growth net of efficiency, tracked in dollars as well as tokens, from serving-price, volume, and platform-revenue data.

\end{abstract}
\section*{Contributions}
This paper makes eight contributions. First, it formulates inference economics in a bandwidth-denominated unit, dollars per petabyte delivered (\$/PB), given in closed form with an explicit scope qualification for workload shape and utilization; the unit is model-agnostic for bandwidth-bound decode and cleanly separates hardware economics from model choice (Section 2). Second, it establishes the depreciation-conveyor result: the cost gap between entrants and incumbents never closes within the horizon, because amortization continuously delivers newly cheap fleets to whoever bought last cycle's hardware, so the advantage rotates among incumbents rather than transferring to entrants (Section 3). Third, it derives a U-shaped vintage-breakeven curve and shows that the identity of the impaired party is selected by the pricing regime --- coupled pricing breaks the 2028--2029 vintages, sticky pricing breaks the 2026 vintage, and only the 2027 vintage is robust in all cases (Section 6). Fourth, it derives a solvency corridor for the aggregate buildout --- approximately 2$\times$ annual token-demand growth sustained for four years, with the threshold moving between 1.6$\times$ and 2.4$\times$ across efficiency-trend bounds, and with AI for Science and Industrial Innovations demand identified as the corridor's mission-funded, rationing-proof floor (Section 7). Fifth, it conducts a measurement critique of public token-demand trackers, identifying five systematic upward biases and re-denominating the corridor in dollars (Section 9.1). Sixth, it identifies a projection-vintage problem: all optimistic demand projections predate the Q2 2026 regime break from token maximization to token minimization, and are therefore upper bounds (Section 9.3). Seventh, it analyzes the Chinese full-stack decoupling (domestic HBM, standard ISA, open models) and the Western circular-financing web as a coupled transmission mechanism for the crash scenario, with the telecommunications precedent as the structural overlay and the CLOUD Act's jurisdictional reach as the reason sovereignty requires domestically owned and operated infrastructure rather than open weights alone (Section 10.1). Eighth, it derives success, mediocrity, and loss probabilities for a greenfield entrant shipping custom silicon into a new datacenter, showing that custom silicon removes the merchant margin but not the memory premium, bounding the achievable advantage; the resulting central outcome distribution is 25/34/41 (success/mediocre/loss), managed by a staged go/no-go procedure (Section 10.2).

\section{Market Context, Mid-2026}
The memory market is in the most severe repricing in the industry's history. Conventional DRAM contract prices rose on the order of 90\% in Q1 2026 over Q4 2025 [1] and continued rising through Q2, with the three suppliers reallocating the large majority of wafer capacity toward HBM and server products; new fab capacity arrives meaningfully only in 2027--2028, and SK hynix leadership has warned the shortage could persist to 2030 [2]. Memory now constitutes roughly 40--50\% of accelerator bill-of-materials and a rising share of total data-center capital cost.

Simultaneously, the open-weight frontier has effectively caught up for the capability band that serves the large majority of workloads. GLM-5.2 (744B total parameters, ~40B active, MIT license, June 2026) [3] matches or exceeds proprietary flagships on long-horizon coding and agentic benchmarks at roughly one-sixth the serving price, and aggressive quantization places it on workstation-class unified-memory hardware. Inference efficiency is improving on several independent axes at once: KV-cache compression approaching the information-theoretic bound [4], MoE sparsity decoupling capability from per-token compute, and routing systems that direct the majority of tokens to the cheapest adequate model.

On the infrastructure side, Meta is standing up a compute-resale business (``Meta Compute'') [5] on top of a capex program guided to \$125--145B for 2026, and xAI/SpaceX has leased the entirety of Colossus 1 --- roughly 300 MW and over 200,000 accelerators --- to Anthropic at approximately \$1.25B per month [6], with further leases signed to Google and Reflection AI. The neocloud sector repriced sharply downward on these announcements. OpenAI's Stargate program, by contrast, represents the largest concentration of memory purchased at or near peak prices, with reports attributing to it up to 40\% of global DRAM output [7].

\section{Analytical Framework and Core Assumptions}
\subsection{Terminology and definitions}
Because several near-synonyms carry distinct technical meanings in this paper, we fix them here. A \textbf{vintage} is a purchase-year cohort of capacity; its economics evolve deterministically over its life --- full cost while amortizing, marginal cost thereafter --- so vintage is a property of \textit{when hardware was bought}. A \textbf{legacy fleet} is a vintage that has completed, or substantially completed, depreciation and therefore operates at marginal cost; every vintage becomes legacy in time, so the term describes hardware age, not merit and not ownership. An \textbf{incumbent} is a market-position term: an operator holding prior vintages whose capital is sunk. Incumbency is ownership of the depreciation conveyor's output; a firm can simultaneously be an incumbent in capacity and a newcomer in a product market --- Meta entering compute resale is the canonical example. An \textbf{entrant} (or \textbf{newcomer}) is an operator that must acquire the current vintage at full cost; the distinction from incumbents is one of balance sheet, not brand, age, or technical skill. The \textbf{incumbent floor} is the lowest marginal cost among fielded legacy fleets --- the price level incumbents can profitably defend indefinitely and entrants cannot reach for the life of their hardware.

Cost terms follow standard usage: \textbf{full cost} includes capital amortization; \textbf{marginal cost} excludes it and is the economically relevant figure once capital is \textbf{sunk}. On the demand side, the \textbf{premium tier} comprises tokens commanding frontier-model pricing on grounds of capability, reliability, or compliance, and the \textbf{mass tier} comprises tokens served by good-enough models at floor-anchored prices. The pricing-regime axis has two poles, used throughout as counter-terminology: under \textbf{sticky pricing}, premium prices hold in absolute dollars per token, defended by contracts, compliance lock-in, and switching costs; under \textbf{coupled pricing}, routing arbitrage drags premium prices down in roughly fixed ratio to collapsing mass prices. \textbf{Metered inference} is inference billed by an infrastructure or API provider and is the only demand that services capex; \textbf{total inference} additionally includes on-premises and owned-hardware inference invisible to providers --- the wedge between the two is the reallocation channel of Section 9.2. The \textbf{solvency corridor} is the region of (demand growth, efficiency trend) space in which announced capacity earns its cost of capital, and the \textbf{regime break} denotes the Q2 2026 doctrinal inversion from token maximization to token minimization that separates projection vintages.

Decode-phase inference is bandwidth-bound: throughput is proportional to delivered HBM bandwidth regardless of model identity, because every generated token requires streaming the resident working set (active weights plus KV cache) from memory. The natural unit of inference cost is therefore dollars per petabyte moved (\$/PB), which is model-agnostic and cleanly separates hardware economics from model choice. Full cost comprises straight-line four-year amortization of accelerator price times a 1.5$\times$ system multiplier, plus power at \$0.08/kWh and PUE 1.3, plus \$0.25/GPU-hr allocated operations. Marginal cost omits amortization and is the economically relevant floor for fleets whose capital is sunk. Memory-bandwidth utilization (MBU) is taken as 0.50 for Hopper, 0.55 for Blackwell, and 0.60--0.62 for Rubin-generation parts. Platform assumptions: H100 3.35 TB/s at 700 W; H200 4.8 TB/s at 700 W; B200 8 TB/s at 1,000 W; GB300 8 TB/s at 1,400 W and 288 GB; Rubin 16 TB/s HBM4 at 1,800 W; a 2029 refresh at 20 TB/s and 2,200 W. Purchase prices follow two HBM branches: a base branch in which memory prices normalize from 2028, and a shortage branch in which elevated pricing persists through 2030. All parameters are explicit and the model is intended for iteration rather than false precision; conclusions below are robust to $\pm$20\% parameter perturbation except where noted.

Formally, for a fleet unit with accelerator price $P\_{\mathrm{acc}}$, system multiplier k, straight-line depreciation life T years, electrical draw W watts at power-usage effectiveness PUE and electricity price $c\_e$, allocated operations cost o per accelerator-hour, theoretical memory bandwidth B in TB/s, and memory-bandwidth utilization $\eta$:

\begin{equation}\label{eq:costpb}
\mathrm{Cost}_{\mathrm{PB}} \;=\; \frac{\dfrac{P_{\mathrm{acc}}\,k}{8760\,T} \;+\; \dfrac{W}{1000}\cdot \mathrm{PUE}\cdot c_{e} \;+\; o}{\,B\,\eta \times 3.6\,} \qquad \left[\$/\mathrm{PB}\right]
\end{equation}
in dollars per petabyte delivered, where B$\cdot$$\eta$$\cdot$3.6 is petabytes moved per hour. The marginal-cost variant drops the first (amortization) term once capital is sunk. Every fleet number in this paper is an instance of this expression evaluated with the Appendix A parameters.

One scope qualification is required before use. \$/PB is the right lower-level unit for saturated, bandwidth-bound decode, and it is used here to compare hardware vintages; it is not a claim that all deployments achieve the same token economics. Realized dollars per million tokens depend additionally on prefill/decode ratio, context-length distribution, batchability, latency objectives, offered load, utilization, KV residency, and traffic burstiness --- a recent serving-cost methodology study finds effective cost on identical H100 hardware ranging from \$0.21 to \$15.25 per million output tokens as a function of utilization and offered request rate [25]. The same qualification cuts against naïve on-premises economics later in the paper: self-hosting is not automatically cheaper unless utilization is high or the hardware is justified by sovereignty and privacy rather than cost.

A final classification concerns capacity that belongs to neither balance-sheet class defined in Section 2.1: sovereign and institutional systems --- national-laboratory and agency machines such as NERSC's Doudna in the United States, the LineShine system analyzed in Section 10.1, and Japan's Fugaku lineage --- procured at negotiated prices under mission budgets rather than market solvency constraints. Such capacity does not participate in the corridor's solvency logic, because it is not financed against token revenue; but it does participate in supply. Sovereign fleets serving open-weight inference add to the mass tier's effective capacity at prices unconstrained by cost recovery, deepening the incumbent floor within the jurisdictions where they operate, and their procurement decisions are among the few demand anchors insulated from the pricing regimes of Section 6.

\section{Incumbent Floor versus Entrant Economics}
Figure 1 shows the trajectory. A new entrant purchasing GB300-class capacity in 2026 pays peak memory prices and faces a full cost near \$0.174/PB, against an incumbent floor of roughly \$0.054/PB set by depreciated H100 fleets whose capital is sunk --- a 3.2$\times$ disadvantage that no operating skill can overcome. The 2027 Rubin generation improves the entrant position to a 1.9--2.0$\times$ gap, the best entry point in the period. Thereafter the gap re-widens: by 2029 the floor falls to \$0.022/PB as B200 fleets complete amortization, while even normalized-HBM new builds cost \$0.072/PB. This is the report's first structural conclusion: the ``memory bought cheap'' advantage is not a one-time inventory coup that expires when H100s obsolesce; it is a rolling property of the depreciation conveyor, and it accrues perpetually to whoever bought last cycle's hardware --- that is, to incumbents as a class. An entrant can only join the class by surviving one full depreciation cycle at negative spread, which is precisely what incumbent limit pricing can prevent: an incumbent pricing tokens anywhere above its own marginal cost but below the entrant's full cost earns profit while guaranteeing the entrant losses for the life of the hardware.

Figure 2 shows the countervailing force. Rubin-generation parts deliver roughly 2.2$\times$ the bandwidth per watt of H100. In power-constrained sites --- which is to say nearly all sites --- the opportunity cost of a power slot eventually dominates. Formally, with delivered bandwidth per watt $\beta$ = B$\cdot$$\eta$/W and market token price p (in \$/PB), retirement of a legacy part occupying a power-limited slot is rational when

\begin{equation}\label{eq:slot}
\beta_{\mathrm{new}}\!\left(p - c^{\mathrm{full}}_{\mathrm{new}}\right) \;>\; \beta_{\mathrm{leg}}\!\left(p - c^{\mathrm{marg}}_{\mathrm{leg}}\right),
\qquad \beta \equiv \frac{B\,\eta}{W}
\end{equation}
that is, when the new part generates more margin per watt, not merely more bandwidth per dollar. Solving this condition with the Appendix A parameters, replacement of H100 by normalized-price Rubin becomes margin-positive whenever market token prices exceed roughly \$0.11/PB, barely above Rubin's own full cost; this inequality is also the mathematical bridge between the hardware economics of this section and the fleet-utilization corridor of Figure 6. Incumbents therefore refresh on schedule in the base branch; in the shortage branch elevated hardware prices delay their refresh too, keeping legacy fleets in service longer and anchoring mass-market prices lower for longer. The shortage branch is thus doubly hostile to entrants: their hardware costs more and the price umbrella above them sits lower.

\begin{figure}[t]\centering
\includegraphics[width=0.92\linewidth]{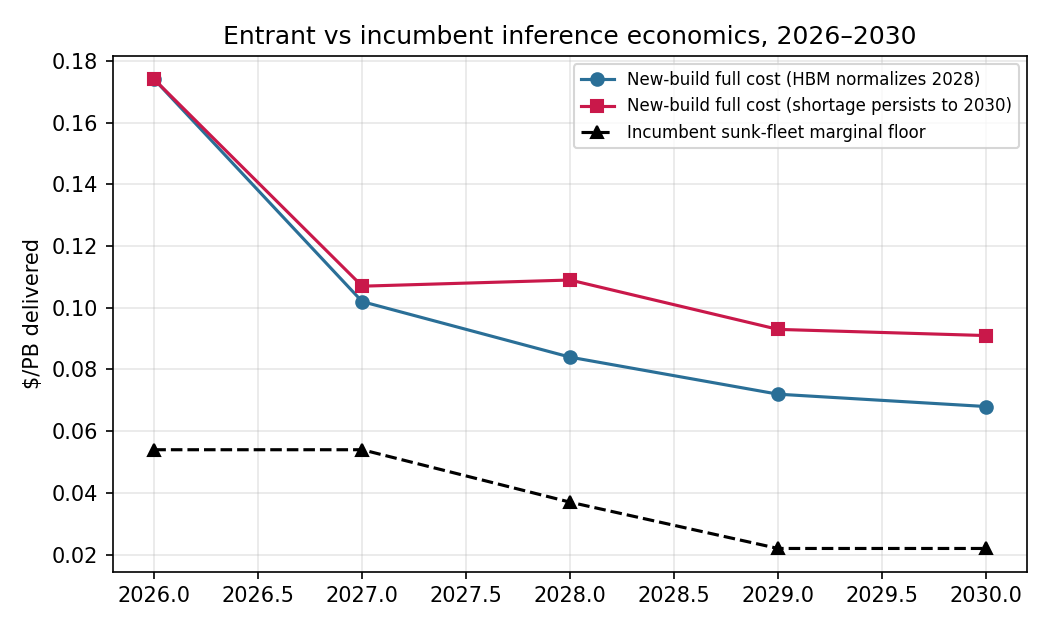}
\caption{New-build full cost vs incumbent sunk-fleet marginal floor, \$/PB delivered, under two HBM branches. The gap narrows to 1.9$\times$ in 2027 and re-widens to ~3--4$\times$ by 2029--30.}\label{fig:1}
\end{figure}
\begin{figure}[t]\centering
\includegraphics[width=0.92\linewidth]{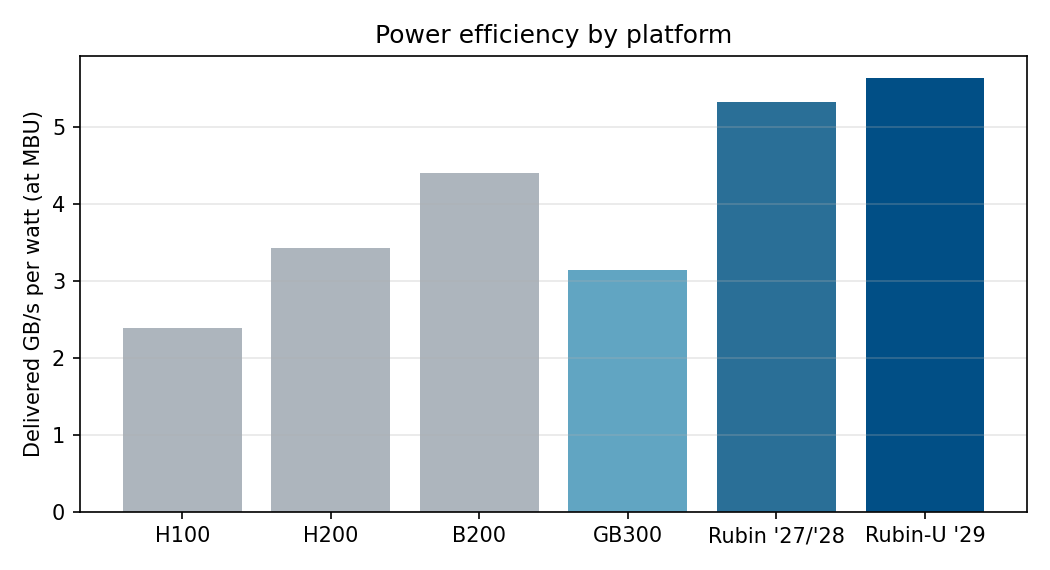}
\caption{Delivered bandwidth per watt. Rubin's ~2.2$\times$ advantage over H100 governs the power-slot displacement condition that ends each vintage's service life in power-limited sites.}\label{fig:2}
\end{figure}
\section{The Capacity--Bandwidth Trade-off}
Unoptimized MoE serving keeps all expert weights and the KV cache resident in HBM: a GLM-5.2-class model at FP8 requires on the order of 750 GB of residency before KV, i.e., roughly ten H100s but only three GB300s per replica. Legacy fleets are therefore bandwidth-cheap but capacity-poor, and serving frontier-scale MoE on them multiplies interconnect traffic and degrades effective MBU. The arbitrage is consequently workload-dependent: depreciated fleets dominate the economics of models up to roughly 100--200 GB residency --- which, after quantization, includes most ``good enough'' open models --- while very large MoE serving favors capacity-rich new silicon. This segmentation cuts in both directions. It preserves a defensible niche for new hardware at the premium end, and it creates a powerful incentive to apply the compression stack (low-bit weights, near-optimal KV quantization, expert offloading) precisely in order to pull large models back onto cheap capacity-poor fleets. Every success of that stack transfers workload from the premium niche to the incumbent floor.

\section{Training-Cost Divergence and the Luxury Segmentation}
Frontier training ambition continues to scale at roughly 2.5$\times$ per year in compute, against hardware cost-per-FLOP improving only ~25\% per year in the base branch and ~10\% per year in the shortage branch, because memory now dominates accelerator cost and memory is not on a Moore-like trajectory. The cost of a frontier-class run therefore rises from roughly \$1.5B in 2026 to \$18B (base) or \$38B (shortage) by 2030 --- Figure 3. Meanwhile the cost of reaching previous-frontier, ``good enough'' capability by reinforcement-learning post-training and distillation on open bases falls from tens of millions of dollars toward single-digit millions, driven by recipe efficiency and the free-rider benefit of each newly released open base. The ratio between the two widens from roughly 40$\times$ today to three-to-four orders of magnitude by 2030.

The economic consequence is the luxury-car structure: frontier closed models remain technically superior and can remain profitable, but only as premium products whose development cost is recovered from a minority segment that genuinely requires frontier capability --- deep agentic reliability, novel science and engineering, regulated high-stakes deployment. The mass token market --- routine coding, retrieval, drafting, routine agents --- is served at a per-token cost the luxury tier cannot approach, by open models whose development cost is three orders of magnitude lower and whose serving runs on the incumbent floor. A closed lab's viable strategies reduce to two: price the premium tier on sticky absolute terms decoupled from the mass market, or descend into the mass market and accept commodity margins on infrastructure it does not own at the floor. The scale mismatch is the crux: if aggregate infrastructure is being financed as if premium demand were mass-sized, the financing assumptions and the market structure are inconsistent, and one of them must break.

\begin{figure}[t]\centering
\includegraphics[width=0.92\linewidth]{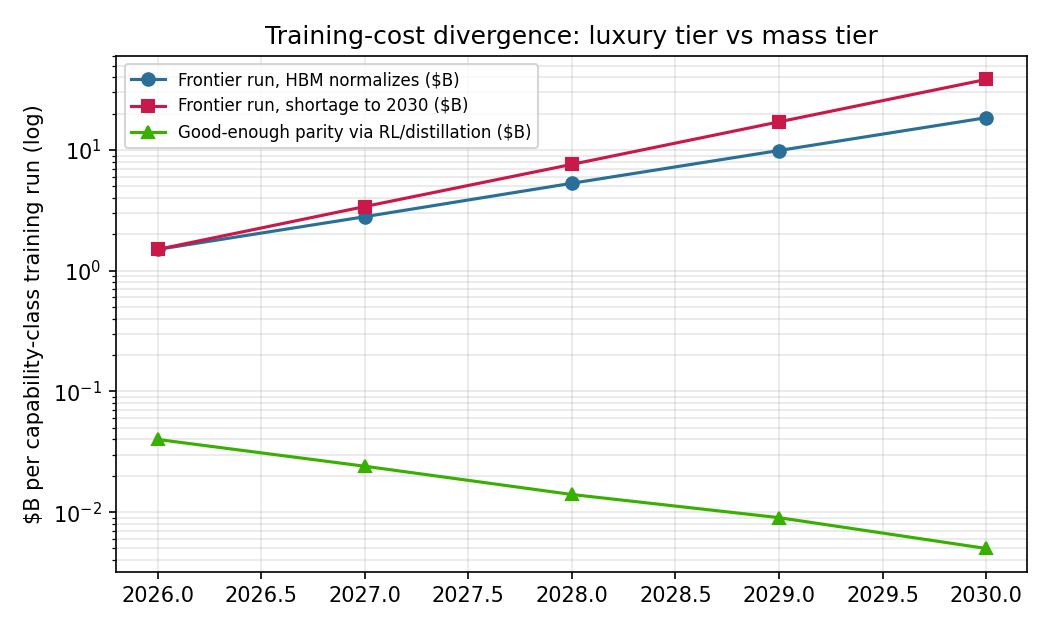}
\caption{Training-cost divergence, log scale. Frontier runs reach \$18--38B by 2030 while good-enough replication falls toward \$5M --- a 3,600--7,400$\times$ gap.}\label{fig:3}
\end{figure}
\section{Vintage Breakeven and Pricing-Regime Sensitivity}
For each purchase vintage we compute the share of served tokens that must earn premium pricing for the fleet to break even, under two pricing regimes. In the coupled regime, routing arbitrage drags premium prices down with the collapsing mass floor (premium fixed at 7$\times$ the prevailing mass price); in the sticky regime, premium pricing holds at \$0.40/PB absolute on the strength of enterprise contracts, compliance lock-in, and agentic reliability that benchmarks understate. Figure 4 shows the result, against a plausible realized premium share of 10--20\% of tokens.

The pattern is U-shaped and regime-dependent. Under coupled pricing the casualties are the late vintages: 2028--2029 purchases require 21--38\% premium share in the shortage branch, well above the plausible band --- this is the failure mode in which routing succeeds and the mass floor sets all prices. Under sticky pricing the late vintages survive comfortably at 10--17\%, and the worst position becomes the 2026 vintage at 31\%, which paid peak hardware prices without a compensating pricing umbrella. The 2027 vintage is robust in every cell at 8--11\%, making it the rational entry year in all futures. The qualitative conclusion is stronger than any single number: someone holding peak-vintage capacity is structurally underwater in every regime; the pricing regime selects who. Routing success breaks the Stargate-class 2028--2029 commitments; routing failure breaks today's peak-price buyers.

\begin{figure}[t]\centering
\includegraphics[width=0.92\linewidth]{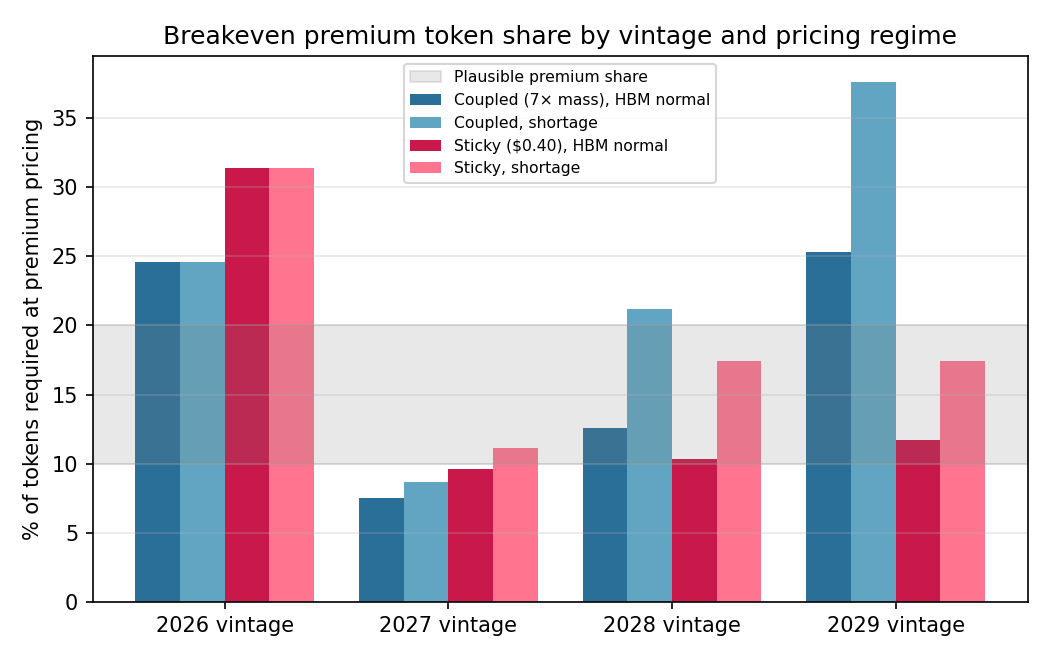}
\caption{Breakeven premium token share by vintage under coupled vs sticky premium pricing and two HBM branches, against a plausible 10--20\% realized premium share (shaded).}\label{fig:4}
\end{figure}
\section{The Demand-Side Solvency Corridor}
The system-level question is whether aggregate demand can absorb the announced buildout at prices that keep new capacity solvent. Installed capacity under announced plans grows roughly 2.6$\times$ by 2029. Demand for delivered bandwidth equals token demand divided by efficiency: bytes moved per token are falling on the order of 30\% per year through KV compression, sparsity, quantization, and routing to smaller models. Figure 5 shows the balance: at 1.5$\times$ annual token growth, fleet utilization collapses to roughly 45\% by 2029 and mass impairment follows; at 2.0$\times$ the system is approximately balanced; at 2.5$\times$ and above, capacity is short throughout and even peak-vintage fleets remain solvent --- the Jevons rescue. The solvency threshold is therefore approximately 2$\times$ annual token growth sustained for four consecutive years. Current agentic token growth runs well above this, but the requirement is persistence, and there is an asymmetry: efficiency gains are cumulative and irreversible, while demand hypergrowth is neither.

\begin{figure}[t]\centering
\includegraphics[width=0.92\linewidth]{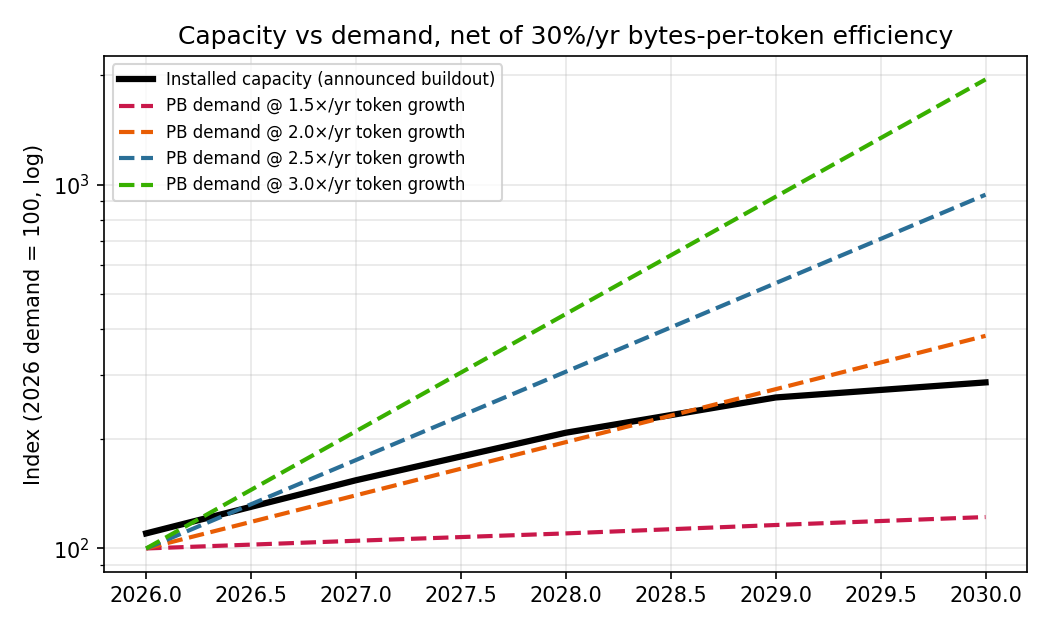}
\caption{Installed capacity vs bandwidth demand at various token growth rates, net of 30\%/yr efficiency. The threshold for system solvency is approximately 2$\times$/yr.}\label{fig:5}
\end{figure}
One demand segment deserves separate treatment in the corridor: AI for Science and Industrial Innovations (AI4SIS). Autonomous scientific agents executing the loop of science --- hypothesis generation, large-scale simulation, and robotic experiment orchestration --- together with their industrial-R\&D counterparts in materials, drug, and engineering design, consume compute that is mission-funded and publicly or strategically budgeted; the US DOE Genesis Mission is the flagship public instantiation. This demand is structurally insulated from the enterprise token-rationing documented in Section 9.1: a national laboratory does not throttle a fusion-design campaign because a CFO's token budget is exhausted, and an industrial R\&D program whose output is patents and products rather than per-seat productivity does not meter itself against SaaS-style budgets. AI4SIS demand is comparatively inelastic, long-horizon, and simulation-heavy --- precisely the profile that keeps bandwidth-hungry fleets utilized through an enterprise demand air-pocket --- and it is the strongest structural argument for retaining a material probability on the Jevons Absorption scenario.

\subsection{Stress test: efficiency deceleration}
The 30\%/yr efficiency assumption is the report's most consequential parameter, and there is a concrete physical argument that it decelerates: KV-cache quantization has reached within sight of the information-theoretic bound --- TurboQuant's error is close to the Shannon limit, meaning the largest single lever is nearly exhausted --- and MoE sparsification faces quality floors. Re-solving the corridor: at 15\%/yr efficiency the solvency threshold falls to about 1.6$\times$ annual token growth; at the baseline 30\%/yr it is 1.9$\times$; if efficiency instead accelerates to 45\%/yr (new levers: speculative decoding at scale, radical sparsity, on-device offload) the threshold rises to 2.4$\times$. Figure 6 maps the full corridor: the 90\%-utilization contour separates solvency from impairment across the (demand growth, efficiency) plane. Efficiency deceleration is the strongest counterargument to the breakdown thesis --- it widens the corridor by roughly a third --- but it carries a second-order sting for closed labs: if efficiency stalls, serving costs stop falling, and the luxury tier's operating expense burden grows with model scale rather than shrinking behind it.

\begin{figure}[t]\centering
\includegraphics[width=0.92\linewidth]{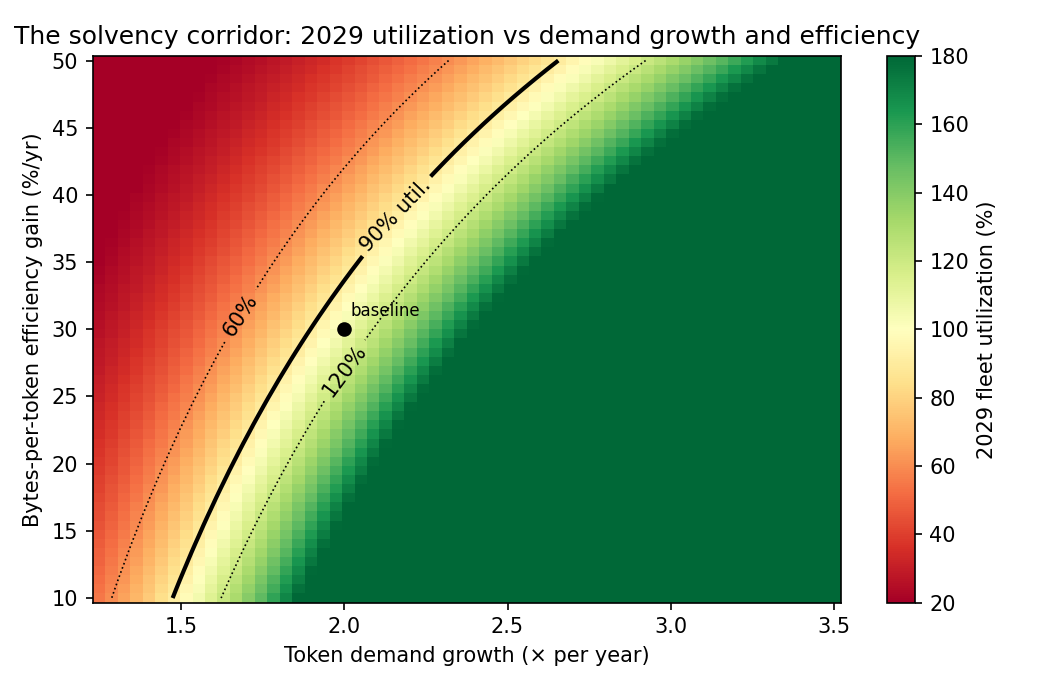}
\caption{The solvency corridor. 2029 fleet utilization across token-demand growth and efficiency-gain rates; the 90\% contour divides solvency from impairment. Baseline marked at (2.0$\times$, 30\%).}\label{fig:6}
\end{figure}
\section{Scenarios, 2026--2030, with Revised Probabilities}
The corridor logic reorganizes the original scenario set. Each scenario is now identified with a region of the (token-demand growth, premium-pricing stickiness, HBM branch) space, and probabilities are revised accordingly.

\begin{table}[t]\centering\footnotesize
\caption{Scenario definitions and revised probabilities (fifth-round assessment).}
\begin{tabular}{L{2.8cm}L{0.8cm}L{10.2cm}}
\toprule
\textbf{Scenario} & \textbf{P} & \textbf{Defining conditions and consequences} \\
\midrule
S1. Rotating Landlord<br/>Oligopoly & 25\% & Demand near threshold; shortage branch extends legacy fleets. Meta/xAI-class incumbents set mass prices from the sunk-fleet floor; the floor advantage rotates by vintage and never reaches entrants; neoclouds compress or exit; closed labs rent rather than build (the Anthropic--Colossus pattern generalizes). \\
S2. Jevons<br/>Absorption & 20\% & Token growth $\geq$2.5$\times$/yr sustained. Agentic workloads absorb the buildout; all vintages solvent; entrant gap persists but matters less than time-to-power; closed labs retain durable premium on long-horizon reliability. Historically the default outcome for compute markets, with AI4SIS workloads (Section 7) as the inelastic base load. \\
S3. System-Layer<br/>Re-differentiation & 18\% & Sticky premium pricing regime dominates. Weights commoditize; value migrates to orchestration, proprietary RL environments, interaction data, and trust; router-plus-escalation is the margin pool and closed labs are the escalation tier; infrastructure earns utility margins. \\
S4. Commoditization<br/>Crash & 25\% & Token growth <1.7$\times$/yr and/or coupled pricing. Routing plus local inference hollow out centralized demand as 2027--2028 fab capacity lands; impairments concentrate on peak-vintage holders --- 2026 buyers under sticky pricing, 2028--2029 (Stargate-class) commitments under coupled pricing; a major lab recapitalizes or consolidates; memory glut follows in the classic DRAM bust pattern. \\
S5. Geopolitical<br/>Bifurcation & 12\% & Export controls extend to model weights bidirectionally (the Fable-5 restriction as precedent). Western enclave restores closed-lab pricing power; ROW standardizes on Chinese open weights; sovereign open-weight inference capacity --- including Japan's --- appreciates sharply in strategic value. \\
\bottomrule
\end{tabular}\label{tab:1}
\end{table}
Across five revision rounds, the Crash scenario rose from an initial 15\% to a peak of 30\% before being tempered to 25\% on external-review grounds --- the post-break evidence window, at weeks, is too short to reset a base case: first on the structural finding that the entrant gap never closes and late-vintage breakeven is demanding under coupled pricing, and second on the measurement critique of Section 9.1, which reduced quality-adjusted demand growth to the threshold band with a decelerating second derivative. S1 has been redefined from fixed incumbency to a rotating oligopoly per the depreciation-conveyor result. The present assessment is that the industry has moved from a growth default to a corridor regime: crash risk is unusually high and co-modal with the landlord outcome, the modal structure is a contested landlord--orchestration market, and Jevons absorption remains a live upside path. Bifurcation, at 12\%, is better read as a cross-cutting overlay than as a mutually exclusive world --- the June 2026 export-control episode [27] shows that policy constraint of frontier access can occur inside S1, S3, or S4 rather than replacing them. The financing of the buildout remains concentrated in the node most exposed across the downside cells.

\section{Survey: Tracking the Corridor Variables in the Data}
The corridor's coordinates --- token-demand growth and efficiency --- are observable today. On volume, the public record is unusually rich. Google disclosed a trajectory from roughly 9.7 trillion tokens per month in May 2024 to 480 trillion at I/O 2025, 1.3 quadrillion by October 2025, and 3.2 quadrillion by May 2026 [12] --- a stated 7$\times$ year-over-year growth, with the most recent seven months annualizing near 4.7$\times$. Microsoft reported Foundry customer token processing accelerating roughly 30\% quarter-over-quarter in fiscal Q3 2026 [13] ($\approx$2.9$\times$ annualized). OpenRouter, the most useful neutral multi-vendor sample, grew from about 5 trillion tokens per week in spring 2025 to over 31 trillion by late May 2026 [11] ($\approx$4--6$\times$ year-over-year, accelerating intra-2026), with agentic traffic overtaking human traffic around February 1, 2026 and consuming roughly fifteen times the tokens per request. Third-party trackers place China's aggregate consumption near 180 trillion tokens per day by February 2026 [14], up from about 30 trillion in mid-2025 --- six-fold in eight months --- with video generation (~350 thousand tokens per ten-second clip) opening an additional demand regime. Goldman Sachs projects 24$\times$ growth in token consumption between 2026 and 2030 [8], i.e., a compound rate of about 2.2$\times$ per year --- notably, almost exactly the solvency threshold derived independently in Section 7. Figure 7 assembles the trackers.

On prices and efficiency, the fixed-capability price level is collapsing --- Stanford HAI finds GPT-3.5-class inference roughly 280$\times$ cheaper over two years [9], and a16z estimates about 10$\times$ per year at constant capability [10] --- but this conflates hardware efficiency, model right-sizing, routing, and margin compression; the hardware-level bytes-per-token trend consistent with these observations is the 30--45\%/yr band used in Section 7. On the pricing-regime question, the current evidence favors stickiness at the top: on OpenRouter in late May 2026, Anthropic captured roughly 42\% of platform revenue on about 11\% of token share, with Opus-class output priced near \$25--30 per million tokens against \$0.09--0.18 for DeepSeek V4 Flash --- a two-order-of-magnitude spread that has so far not coupled downward. Meanwhile the mass tier races down: models under \$1 per million input tokens reportedly grew from ~18\% of OpenRouter requests in January 2026 to ~41\% by June, Chinese open-weight models crossed 50\% of platform token share in early 2026 (from under 2\% in late 2024), and hyperscalers are institutionalizing routing (Bedrock's Intelligent Prompt Routing across 110+ model variants; equivalent machinery in Vertex and Foundry) --- which, as hyperscaler operating margins above 33\% on open-weight tokens demonstrate, transfers the model-provider margin to the infrastructure layer rather than destroying it. Two counter-signals warrant close tracking: enterprise budget rationing has begun (firms exhausting annual AI budgets in months [24], internal license cancellations, metering initiatives, and survey evidence that roughly 60\% of enterprises report throttling AI spend [26]), which is the first concrete demand-deceleration mechanism observed; and the abrupt June 2026 suspension of a newly released frontier model under a US government directive days after launch demonstrated that premium supply can be policy-fragile, a mechanism belonging to the Bifurcation scenario.

\begin{figure}[t]\centering
\includegraphics[width=0.92\linewidth]{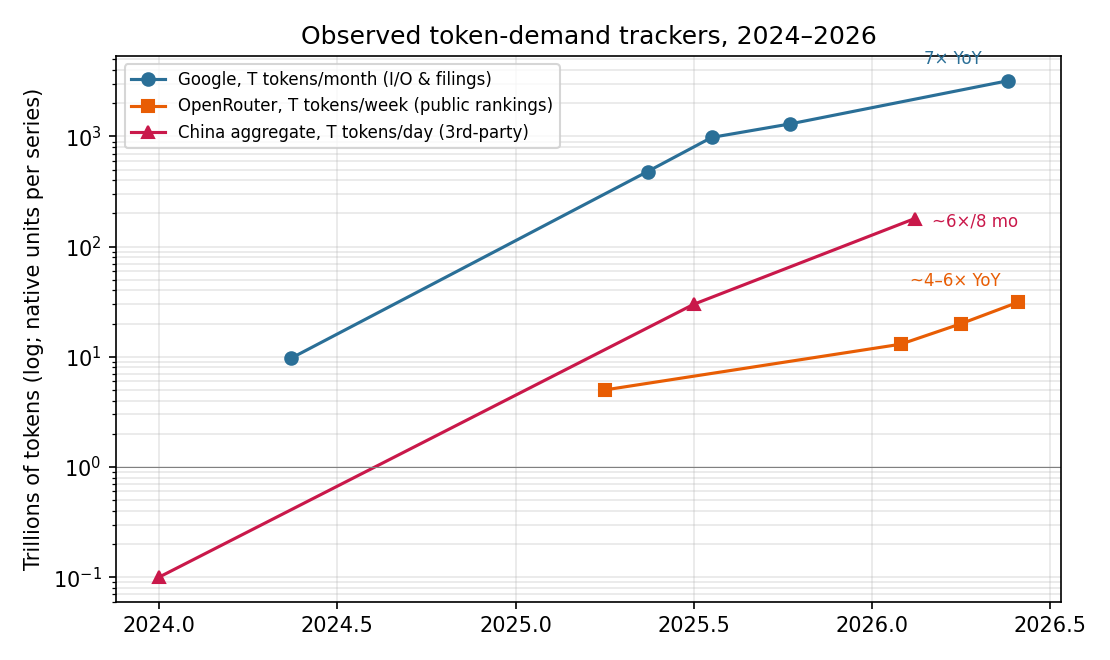}
\caption{Observed token-demand trackers (log scale, native units per series). All primary sources are vendor-reported or third-party aggregations; levels are not directly comparable across series but growth rates are.}\label{fig:7}
\end{figure}
Reading the trackers against the corridor at face value: gross token growth of 4--7$\times$/yr net of a 30--45\%/yr efficiency trend implies delivered-bandwidth demand growing roughly 2.5--5$\times$/yr as of mid-2026 --- above the ~1.6--2.4$\times$ threshold band, consistent with the observed accelerator and memory shortage and with fleets of every vintage currently earning. That is the gross reading; the following subsection examines how much of it survives scrutiny.

\subsection{Measurement critique and revised weights}
The headline trackers systematically flatter underlying demand, for five identifiable reasons. First, supply-injected volume: the largest single tracker aggregates inference that the platform operator pushes onto users --- AI summaries inserted into search results and productivity surfaces at the operator's discretion, plus multimodal processing whose token counts are enormous by construction --- which measures product strategy, not willingness to pay. Second, substitution: router-platform growth includes workloads migrating onto the router from direct APIs, so platform growth exceeds market growth by the migration rate. Third, subsidy: a meaningful share of reported volume, particularly in the Chinese price war, is quantity demanded at below-cost prices and would not survive normalization. Fourth, quality composition: the ~15$\times$ agentic token multiplier is substantially redundant context re-reading --- the lowest-value tokens in the economy and the first to be eliminated as token optimization becomes an enterprise discipline; the budget-rationing behavior now documented across large enterprises is better read as the leading edge of that optimization than as a peripheral counter-signal, since 2027 budgets will be set with full knowledge of the 2026 burn. Fifth, base effects and disclosure selection: the agentic era began in earnest only in February 2026, so mid-2026 growth rates annualize the steepest point of an S-curve, and the visible sample is biased upward because operators publish token metrics when, and only when, they are spectacular.

Two further corrections follow. The infrastructure entries of Meta and xAI, treated in Section 1 as evidence of incumbent strength, are dual signals: capacity offered to the market is also capacity the largest internal AI operators concluded they did not need for themselves --- insiders selling forward is what a marked-down internal demand forecast looks like. And the corridor itself should be denominated in dollars, not tokens: solvency depends on petabytes moved times achievable dollars per petabyte, and if growth concentrates in near-zero-margin open-weight tokens --- the neutral router's blended realization is on the order of \$1 per million tokens against flagship closed pricing near \$25--30 --- bandwidth demand can grow while revenue per unit of capacity stalls. Quality-adjusting for all of the above, underlying demand growth is plausibly 2--3$\times$/yr rather than 4--7$\times$, sitting at or modestly above the threshold band with a visibly decelerating second derivative rather than comfortably above it. The scenario weights are revised accordingly, and the downside projection of Figure 8 --- absolute bandwidth demand peaking around 2028 --- should be treated as at least as likely as the base case rather than as a tail.

\subsection{Demand reallocation: token minimization, privatization, and the sovereignty channel}
Within a span of months in the first half of 2026, the industry's operating doctrine visibly inverted from token maximization --- the agentic everything-in-context style that defined 2025 --- to token minimization as an enterprise discipline, with budget metering, context pruning, and routing-down institutionalized. Concurrently, the requirement to privatize institutional knowledge is pushing deployment on-premises: capable open-weight models running on private data, on hardware from single DGX Spark-class boxes [23] and Mac-mini clusters to rack-scale private inference, with an entire genre of publicly available documentation and demonstration content lowering the operational barrier month by month. On neutral routers, Chinese open-weight models now carry roughly 60\% of token traffic and rising. The analytical consequence is a distinction the corridor model must respect: metered cloud token demand --- the quantity that services hyperscaler and neocloud revenue and therefore capex solvency --- can decelerate even while total inference grows, because a growing share of inference migrates onto customer-owned hardware where it is invisible to, and unmonetizable by, the infrastructure providers whose fleets the buildout is financing. Demand relocation reads as demand destruction on the balance sheets that matter.

The sovereignty channel amplifies this outside the United States. Japanese enterprise AI adoption runs well below US levels [22], with reporting attributing the gap substantially to concern that proprietary data will be exploited by US model providers; the same posture recurs across much of Europe and Asia. The consequence is not that this demand never materializes --- it is that when it does, the marginal non-US enterprise adopter enters directly at the on-premises open-weight stage, skipping the metered cloud-API stage entirely. The largest reservoir of untapped enterprise demand is thus structurally routed toward open models on private or sovereign infrastructure, and away from the revenue lines of OpenAI, Anthropic, and the hyperscale fleets built to serve them. The automotive analogy is exact in its mechanism: the premium incumbents' technology lead did not prevent cost-structure-driven mass-market capture by Chinese manufacturers in electric vehicles, and models are more substitutable than cars --- a routing table switches suppliers in days, with no dealer network, no charging standard, and no resale value defending the incumbent. The one respect in which the analogy understates the incumbents' position is that they can respond instantaneously in kind --- by price, or by releasing weights --- which vehicles never could; whether their balance-sheet commitments permit that response is precisely the question of Section 10.1.

\subsection{Projection vintage and the regime break}
A final correction concerns not the measurements but their dates. Every optimistic projection cited in this survey is a product of the maximization regime: the Goldman 24$\times$ forecast was published in early May 2026, the tracker growth rates run through May, and the agentic multipliers were measured at the height of everything-in-context engineering. If the doctrinal inversion to token minimization occurred across the client industry in the second quarter of 2026 --- as budget-rationing behavior, metering rollouts, and the collapse of average tokens-per-request among enterprise deployments indicate --- then all trailing-CAGR extrapolations are measurements of a regime that no longer exists, and extrapolating them across the break is a category error. The methodological consequence is that pre-break projections, including this report's own quality-adjusted 2--3$\times$ estimate, should be treated as upper bounds; the go-forward prior must be re-anchored exclusively on post-break data, which as of this writing spans only weeks. Accordingly: pre-break projections, the Goldman-path case of Figure 8 included, are treated as optimistic bounds until two quarters of post-break data establish the new slope, and the downside path --- bandwidth demand peaking around 2028 --- is elevated from tail risk to co-equal stress case alongside the base case, to be promoted or dismissed by the post-break time series rather than by argument.

One decomposition keeps the correction honest. Token minimization is, in this report's framework, an efficiency shock rather than a demand shock: it reduces tokens per task, entering through the efficiency parameter (pushing it toward or beyond the 45\%/yr stress bound and the solvency threshold toward 2.4$\times$ or higher in task terms), while the number of tasks can continue growing independently --- and the one countervailing datum is that neutral-router volumes were still accelerating through late June. Whether minimization is offsetting or overwhelming task growth is therefore directly observable, at weekly cadence, in tokens-per-task and dollars-per-task; the tracker of Section 10 should run monthly rather than quarterly for at least two quarters until the post-break slope is estimated. The scenario weights move on this correction, tempered by the brevity of the evidence window: the Crash rises to parity with the landlord outcome rather than beyond it, and the growth case is retained as a live upside path.

\section{Projections to 2030 and Concrete Scenario Instantiation}
Figure 8 projects delivered-bandwidth demand against the announced buildout under four cases, holding capacity frozen to expose the imbalance each case implies (in reality capacity would chase demand upward in the best cases and FIDs would be cancelled in the worst --- the frozen comparison identifies which pressure operates). In the best case, agentic and video hypergrowth persists with slow decay (6$\times$ tapering to 2$\times$ by 2030, 30\%/yr efficiency): demand exceeds frozen capacity twenty-fold by 2031, the shortage never ends within the horizon, memory pricing stays firm, and every vintage including 2026's peak purchases is rescued by scarcity pricing. In the base case, growth follows the Goldman 24$\times$ path (a constant 2.2$\times$/yr): the system rides the corridor's edge --- utilization near 100\% through 2027, tightening thereafter --- rewarding disciplined capacity addition and punishing nothing except late, expensive vintages in the shortage branch. In the downside case, rationing and routing bite (3$\times$ in 2026 decaying to 1.1$\times$ by 2030, 35\%/yr efficiency): absolute bandwidth demand peaks around 2028 and then declines; utilization falls through 80\% in 2029 and below 60\% in 2030; impairments begin among neoclouds in 2029 and reach peak-vintage hyperscale capacity by 2030. In the worst case (2.5$\times$ fading to $\leq$1.2$\times$ with efficiency accelerating to 40\%/yr as token optimization becomes an enterprise KPI), demand peaks in 2027 and the crash arrives in 2028, compressed by the arrival of new fab supply into falling demand --- the classic DRAM bust with an AI accelerant.

\begin{figure}[t]\centering
\includegraphics[width=0.92\linewidth]{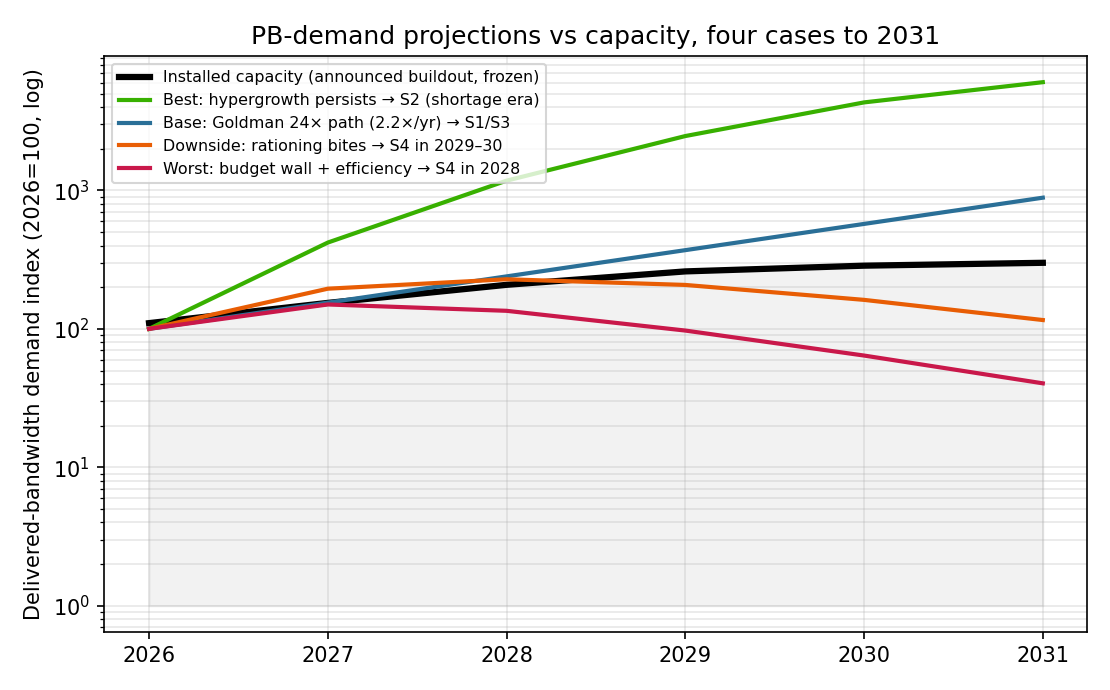}
\caption{Delivered-bandwidth demand under four cases vs frozen announced capacity, 2026--2031, log scale. Case-to-scenario mapping shown in legend.}\label{fig:8}
\end{figure}
\subsection{The full-stack shock and the circular-finance transmission}
Two developments sharpen the downside cases into a concrete transmission mechanism. The first is the demonstration, between April and June 2026, of a complete Chinese AI-compute stack: the LineShine system in Shenzhen --- 40,960 Huawei-designed Armv9 LX2 processors, each pairing on-package domestically produced HBM (roughly 32 GB at ~4 TB/s) with 256 GB of DDR5, SME/SVE matrix units, the domestic LingQi interconnect, domestic storage and operating system --- took the \#1 TOP500 position at 2.198 EFLOPS sustained, the first verified Chinese submission since 2019 [15,16]. On AI-relevant metrics the system is candidly a few years behind: its mixed-precision uplift is 3.6$\times$ against 9--11$\times$ for GPU-accelerated US systems, its 52 GFLOPS/W trails El Capitan, and its HBM is generations behind HBM4. But the strategically decisive element is not position; it is decoupling. Home-grown HBM places the Chinese cost curve outside the global memory-price crisis entirely: while Western buildouts absorb the HBM premium at the center of this report's vintage analysis, the Chinese stack's memory input is priced domestically and directed by the state, immune to the shortage inflating everyone else's capex. Combined with open frontier models that reached parity with flagship closed models within months, and an open ecosystem that compounds compute- and memory-saving innovations faster than closed stacks absorb them, the trajectory is the one familiar from electric vehicles: a few years behind in the premium tier, structurally ahead in cost per unit of mass-market capability, and improving on a steeper slope. The correct reading of LineShine is not the benchmark but the slope --- and, stated precisely: LineShine is not yet evidence of superior AI-inference economics; it is evidence that the sovereign, non-CUDA, HBM-integrated CPU path is real at national scale. Architecturally, the LX2 is also a data point in a live debate: its design --- a matrix-enhanced general-purpose CPU (SME/SVE) fed by an on-package HBM-plus-DDR memory hierarchy, in direct lineage from Fugaku's A64FX --- is precisely the direction argued in the three-author ``Do We Still Need GPUs?'' series by Dongarra, Hoefler, and Matsuoka --- the CACM-submission short paper and its extended empirical companion on matrix-enhanced CPUs [17,18]. LineShine is the first at-scale sovereign instantiation of that direction to lead the TOP500, and its existence means the GPU-free architectural route is no longer hypothetical at the very moment GPU economics are burdened by the memory premium.

The second development is the fragility of the financing web on the Western side. 2026 analyses place more than \$800 billion in circular arrangements [19] among a small cohort: NVIDIA invests in and supplies OpenAI; OpenAI carries commitments reported on the order of \$1.15 trillion --- an aggregation of items of heterogeneous legal strength; see Table 2 --- roughly \$300 billion with Oracle, \$90 billion with AMD, \$38 billion with AWS, plus the \$500 billion Stargate joint venture --- while running losses reported near \$14 billion for 2026; Oracle carries a remaining-performance-obligation backlog around \$523 billion [20], heavily concentrated in OpenAI and financed with rising debt; and the providers buy NVIDIA silicon against that backlog, so each leg books revenue or backlog from the same underlying spend. The NVIDIA--OpenAI \$100 billion equity letter of intent stalled in early 2026 amid scrutiny [21] --- simultaneously a warning and a partial deleveraging. Figure 9 draws the web and its transmission: a premium-demand shortfall of the kind Sections 9.1--9.2 make plausible --- open models capturing the mass tier, on-premises migration removing metered demand, sovereignty channels diverting non-US growth --- flows directly through OpenAI's ability to fund its commitments, into Oracle's backlog quality and debt service, into NVIDIA's order book and the marks on its vendor-equity positions, and finally into a memory market whose expanded 2027--2028 fab capacity would land into the glut. This is the concrete mechanism by which the Commoditization Crash propagates, and it is why both the Crash and Bifurcation weights rise in this revision: the same Chinese full stack that pressures the demand side also gives export-control politics --- already demonstrated against frontier model weights --- a second, hardware-side theater.

A third element converts these from cyclical risks into a structural one: the financing asymmetry between the two blocs. The Western buildout is financed on confidence --- equity valuations, debt markets, and the circular commitments of Figure 9 --- and is therefore pro-cyclical: a demand disappointment tightens the loop, cuts capacity, and forces retrenchment exactly when competition intensifies. The Chinese buildout is financed on state industrial policy and is confidence-insensitive and counter-cyclical: it optimizes for market dominance rather than capital-market value, can price below any level a return-seeking Western competitor can sustain, and can hold that posture for a decade because its capital does not require quarterly reassurance. The precedent is telecommunications: Huawei combined patient state-backed capital, aggressive cost structure, and --- critically --- full compliance with open international standards (3GPP) to entrench itself in 4G and 5G infrastructure across most of the world before the West responded, and the response, when it came, was costly, partial, and late. The AI-era analogue of the standards layer is already in place: the LX2 is standard Armv9, and the open-model software stack (PyTorch-level frameworks, open weights, open inference runtimes) is the 3GPP of this cycle --- there is no CUDA moat anywhere in the mass tier. A compelling Chinese full-stack offering on standard ecosystems, exported at state-supported prices to the very sovereignty-sensitive non-US markets identified in Section 9.2, is the Huawei playbook re-run at the compute layer. This overlay does not reweight the scenarios so much as it ratchets them: in every scenario, including the benign ones, Chinese share gains persist across downturns that force Western capacity to contract, because one side's capacity is a bet and the other's is a plan. The probabilities are therefore left unchanged in this revision, but the Bifurcation scenario should be read as bidirectional in origin --- as likely to be triggered by Western defensive response to entrenchment (the telecom pattern) as by offensive export control.

\begin{figure}[t]\centering
\includegraphics[width=0.92\linewidth]{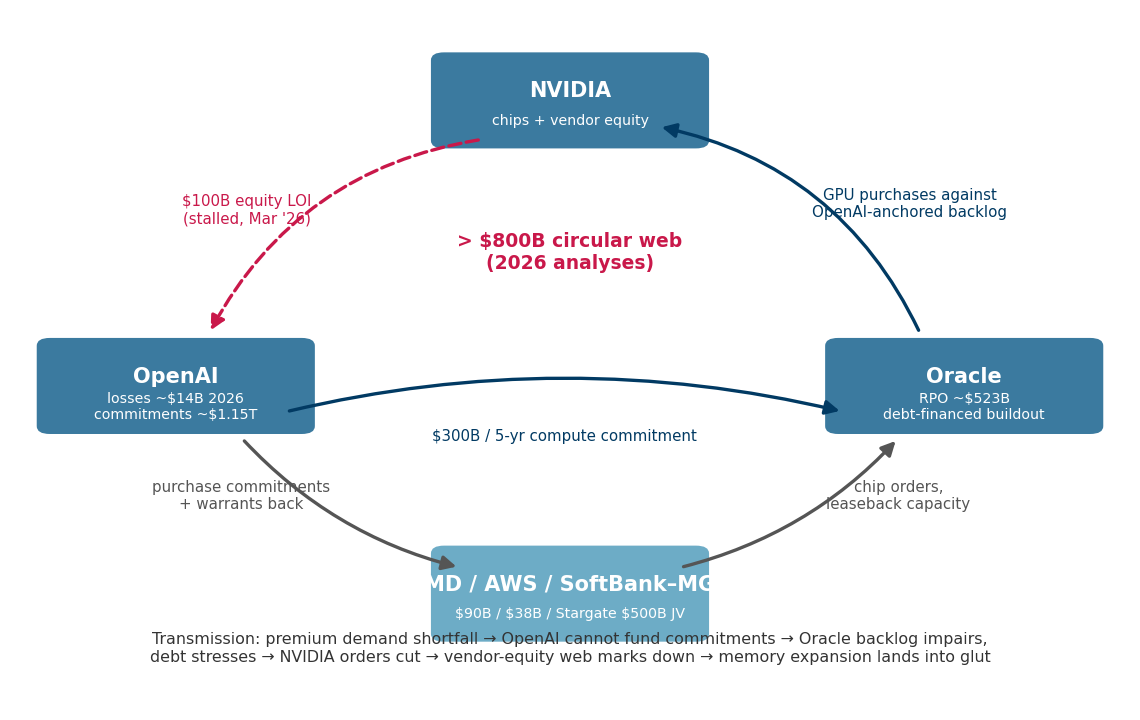}
\caption{The circular-financing web (2026 reported figures; magnitudes approximate and moving) and the crash transmission path. Note that the aggregate mixes contract classes; see Table 2.}\label{fig:9}
\end{figure}
\begin{table}[t]\centering\footnotesize
\caption{Contract-status classification of the financing-web figures: signed or filed items, announced agreements, letters of intent, and analyst aggregations carry very different legal and economic weight. Magnitudes as reported in [7,19,20,21].}
\begin{tabular}{L{5.1cm}L{4.2cm}L{4.9cm}}
\toprule
\textbf{Item} & \textbf{Reported magnitude} & \textbf{Contract-status class} \\
\midrule
Stargate program (OpenAI/SoftBank/Oracle/MGX) & \$500B / 10 GW ambition & Announced program / joint venture \\
OpenAI--Oracle compute agreement & ~\$300B over ~5 yr; +4.5 GW capacity & Reported contract; terms undisclosed \\
OpenAI--AMD agreement & ~\$90B incl. warrant structure & Announced agreement \\
OpenAI--AWS agreement & ~\$38B & Announced agreement \\
Oracle remaining performance obligations & ~\$523B (aggregate backlog) & Audited filing figure; not OpenAI-only \\
NVIDIA--OpenAI equity commitment & \$100B & Letter of intent; stalled February 2026 \\
>\$800B ``circular web'' / \$1.15T totals & aggregates & Analyst and press aggregation of mixed-strength items \\
\bottomrule
\end{tabular}\label{tab:2}
\end{table}
Instantiated concretely, the scenarios read as follows. Jevons Absorption (S2, best case) is the world in which the mid-2026 trackers simply continue --- with AI-for-Science-and-Industrial-Innovations demand compounding on top of enterprise agents as the rationing-proof base load: OpenRouter-class volumes compound, hyperscalers monetize open-weight tokens at infrastructure margins exactly as Bedrock, Vertex, and Foundry are structured to do, the announced OpenAI and Anthropic public offerings complete at or above their contemplated valuations, Stargate's memory commitments are absorbed by scarcity pricing, and the DRAM shortage runs to the end of the decade as SK hynix leadership has warned. The Rotating Landlord Oligopoly (S1, base case) is already visibly assembling: the xAI--Anthropic Colossus lease and Meta Compute are its founding institutions, the neocloud repricing of June 2026 its first casualty list; in this world OpenAI's rational move is to convert Stargate ownership tranches into leases, and the asset-light lab (currently Anthropic, renting depreciating fleets at landlord prices below its alternative cost of ownership) holds the structurally superior position. System-Layer Re-differentiation (S3) is instantiated by the single most striking datum in the survey --- 42\% of router revenue on 11\% of tokens --- generalizing: the escalation tier formalizes, premium pricing holds in absolute terms, and the routers themselves (OpenRouter, Bedrock's routing layer, and their successors) become the market-making institutions of the token economy. The Commoditization Crash (S4, downside and worst cases) fires when two consecutive quarters show platform token growth annualizing below roughly 1.7$\times$ while premium API prices begin cutting toward the mass tier --- watch flagship closed-model price reductions as the coupling tell --- followed in sequence by neocloud covenant breaches, cancellation or lease-conversion of 2028--2029 FIDs, a memory glut as new fab capacity lands into falling demand, and consolidation among frontier labs, with the largest owned-infrastructure balance sheet restructuring first. Geopolitical Bifurcation (S5) already has its mechanism demonstrated --- export controls applied to a newly launched frontier model on June 12, 2026 and lifted June 30 [27] --- and its trigger is the extension of such controls to open weights in either direction; with Chinese models now a majority of neutral-router token share, a Western restriction would forcibly re-shore that volume onto Western models and infrastructure, restoring closed-lab pricing power at the cost of global efficiency, and sharply appreciating the value of sovereign open-weight serving capacity outside both blocs. Jurisdiction compounds the point: under the US CLOUD Act, data held on infrastructure controlled by US providers is reachable by US legal process regardless of where the servers physically sit, so sovereignty in this scenario requires jurisdictional independence of the operator and the infrastructure, not merely open weights or domestic siting --- a requirement that channels the Section 9.2 on-premises migration specifically toward domestically owned and operated capacity.

\subsection{Greenfield entry with custom silicon: success, mediocrity, and loss}
Consider finally an entrant executing the Meta move without Meta's balance sheet: shipping its own custom accelerator and standing up a new datacenter for it --- no sunk fleet, no depreciated vintage, silicon and shell both bought at current prices. Custom silicon attacks exactly one term in the cost stack: the merchant vendor's margin. At Rubin-class specifications (16 TB/s), an owner-cost build --- logic die, packaging, and amortized NRE of several hundred million dollars over ~100k units --- lands near \$42--50k per accelerator against \$58--78k merchant, a 20--35\% capital saving. It does not attack the memory term: HBM, at 45--50\% of bill-of-materials, is priced by the same three suppliers under the same shortage, and a small buyer typically pays more and is allocated later, not less and sooner. Nor does it attack the software term: first-generation custom parts realistically achieve memory-bandwidth utilization of 0.45--0.55 against 0.60 for the merchant ecosystem. Computing through: a well-executed custom build reaches roughly \$0.072/PB, expected execution lands near \$0.082 --- approximately merchant parity --- and a delayed or software-impaired build sits at \$0.105 or worse, while the incumbent floor lies at \$0.022--0.037 throughout. The structural conclusion is sharp: custom silicon can at best neutralize the merchant margin, bounding the achievable advantage over merchant new-builds near 25--30\%, and leaving the entrant 2--4$\times$ above the incumbent floor exactly like every other entrant. The three genuine escape routes are captive anchor demand (a guaranteed internal workload absorbing the fleet, the pattern that made hyperscaler in-house silicon succeed), a memory-side strategy --- secured long-term HBM supply at pre-negotiated prices, or architectural differentiation away from commodity HBM entirely via capacity-tiered or processing-in-memory designs --- and timing into the robust 2027 vintage window rather than the exposed 2028--2029 one.

Table 3 conditions the outcome on the scenario set. \textbf{Success} is defined as risk-adjusted returns above the cost of capital with sustained high utilization and delivered cost at least 15\% below merchant new-builds; \textbf{mediocre} as survival with parity economics or partial utilization, where the venture's value is strategic (an internal cost hedge and supply-chain leverage) rather than financial; \textbf{loss} as impairment or exit. The conditional entries are elicited subjective probabilities --- judgmental assessments conditioned on each scenario's definitions, not model outputs --- and should be read as central estimates within bands: success 15--30\%, mediocre 25--40\%, loss 35--55\%. Under the current scenario weights the central unconditional distribution is approximately 25\% success, 34\% mediocre, and 41\% loss; conditioned on the downside stress case alone (bandwidth demand peaking around 2028), it deteriorates to roughly 10/30/60. The staged gate logic below is more robust than any of these point values. The distribution is strongly controllable by the escape routes: securing all three --- anchor demand, a memory strategy, and 2027 timing --- roughly doubles the success probability toward 45--50\%, while entering without a memory strategy pushes the loss probability above 60\% regardless of silicon quality, because the venture then holds the industry's scarcest input at spot prices with the industry's weakest procurement position.

\begin{table}[t]\centering\footnotesize
\caption{Elicited subjective outcome probabilities (central estimates) for a greenfield custom-silicon entrant with a new datacenter, conditional on scenario; bands of roughly $\pm$7 points apply per cell. Conditionals reflect utilization, pricing regime, and vintage exposure; execution risk held at expected levels.}
\begin{tabular}{L{5.4cm}cccc}
\toprule
\textbf{Scenario} & \textbf{P(scenario)} & \textbf{Success} & \textbf{Mediocre} & \textbf{Loss} \\
\midrule
S1 Rotating Landlord Oligopoly & 25\% & 15\% & 45\% & 40\% \\
S2 Jevons Absorption & 20\% & 65\% & 25\% & 10\% \\
S3 System-Layer Re-differentiation & 18\% & 25\% & 45\% & 30\% \\
S4 Commoditization Crash & 25\% & 2\% & 18\% & 80\% \\
S5 Geopolitical Bifurcation & 12\% & 30\% & 40\% & 30\% \\
Weighted (all scenarios) & 100\% & 25\% & 34\% & 41\% \\
Conditional on downside stress case & --- & ~10\% & ~30\% & ~60\% \\
\bottomrule
\end{tabular}\label{tab:3}
\end{table}
The outcome distribution above is not a verdict to accept but a process to manage, and the correct managerial form is a staged go/no-go procedure whose gates are synchronized to the observables this paper has already defined. The logic of staging follows from an asymmetry: the crash scenario's signatures --- the tokens-growing-dollars-stalling divergence, premium-price coupling, sub-threshold post-break demand slope --- are all observable at monthly cadence, months to quarters before the capital-intensive commitments (datacenter FID, volume silicon, fleet expansion) must be made. A greenfield entrant that commits capital in gate order therefore places the majority of its capital behind the gates at which S4 would already have been detected. Table 4 specifies the procedure: what is measured, the threshold, the calendar point, and the action on each side. Under the current weights, staged entry preserves most of the 25\% unconditional success probability while cutting the loss probability on capital actually deployed from 41\% to roughly 24\%, because gates G2 and G3 --- behind which the datacenter shell, power contracts, and volume fleet sit --- are crossed only in worlds where the crash signatures are absent. The residual ~24\% is irreducible execution and timing risk, not scenario risk.

Three design rules govern the gates. First, every metric is one already in the Section 10 tracker --- the venture requires no private information, only discipline. Second, each no-go branch is a pivot with salvage value, not an abandonment: halted NRE pivots to leasing incumbent capacity (the Anthropic pattern); a shell built lease-ready converts to colocation; a capped fleet serves the anchor tenant as a strategic hedge --- the deliberate-mediocre posture that hyperscaler in-house silicon programs occupied profitably for years. Third, the standing kill criteria apply at every gate without exception, because they mark regime changes that no local execution success can outrun: two consecutive quarters of the crash signature; flagship closed-model price cuts exceeding ~30\% (the onset of coupling); or HBM contract-price re-acceleration above ~20\% quarter-over-quarter without a secured long-term agreement in hand.

\begin{table}[t]\centering\footnotesize
\caption{Staged go/no-go decision procedure for greenfield custom-silicon entry. All metrics are drawn from the standing tracker; no-go branches are pivots with salvage value.}
\begin{tabular}{L{2.4cm}L{1.8cm}L{5.4cm}L{1.8cm}L{2.6cm}}
\toprule
\textbf{Gate} & \textbf{Timing} & \textbf{Metrics and thresholds} & \textbf{Go} & \textbf{No-go action} \\
\midrule
G0 Pre-commitment screen & Before NRE (now) & Anchor-demand LOI $\geq$60\% of fleet-life bandwidth; HBM long-term-agreement term sheet at $\leq$ base-branch prices, or non-HBM architecture; volume silicon in racks $\leq$ Q4 2027 & Commit NRE & Do not enter; lease incumbent capacity \\
G1 Tape-out commitment & ~Q4 2026 & Post-break metered demand slope $\geq$1.8$\times$/yr annualized (3-mo window, task-normalized) AND platform dollar-demand growing; premium/mass price-ratio compression <20\% QoQ & Commit silicon & Halt NRE; pivot to leasing \\
G2 Datacenter FID & ~Q2 2027 & First-silicon MBU $\geq$0.45; anchor LOI converted to contract; merchant 2026-vintage utilization $\geq$85\%; no major announced-FID cancellations & Commit shell + power & Build shell lease-ready/colo; silicon continues for anchor only \\
G3 Volume ramp & Q4 2027--Q1 2028 & Delivered \$/PB $\leq$1.15$\times$ contemporaneous merchant; corridor variables above threshold for 2 consecutive quarters & Expand to merchant scale & Cap fleet at anchor size (hedge posture) \\
Standing kill criteria & Every gate & Crash signature 2 consecutive quarters; flagship price cuts >30\% (coupling onset); HBM contract re-acceleration >20\% QoQ without secured LTA & --- & Freeze all expansion at current gate \\
\bottomrule
\end{tabular}\label{tab:4}
\end{table}
The quarterly tracker this analysis implies is compact, and after Section 9.1 it is denominated in dollars as much as tokens: platform token growth annualized (OpenRouter weekly volumes; Google and Microsoft disclosures) against the 1.6--2.4$\times$ threshold band; platform and hyperscaler AI revenue growth alongside it, with blended realized \$/Mtok as the divergence alarm --- tokens growing while dollars stall is the crash signature; the premium-mass price ratio (flagship closed output pricing vs leading open-weight pricing) as the coupling indicator; premium revenue share on neutral routers (the 42\%-on-11\% datum) as the stickiness indicator; HBM contract pricing against the 2028 normalization assumption; and the fate of announced 2028--2029 FIDs --- proceed, shrink, or convert to lease --- as the market's own revealed forecast.

\section{Implications}
For closed frontier labs, the analysis argues for asset-light serving (leasing incumbent capacity rather than owning peak-vintage hardware), for pricing premium capability on sticky absolute terms decoupled from mass-market drift, and for treating the mass tier as a routing destination rather than a margin source. The lab whose balance sheet embeds mass-sized infrastructure commitments financed at peak memory prices carries the largest single-point exposure in the industry; the lab that rents from the rotating landlords converts the incumbents' arbitrage into its own cost advantage. For infrastructure investors, vintage timing dominates operating skill: 2027 deliveries are robust in every regime examined, while 2026 and 2028--2029 vintages are each fatally exposed to one pricing regime, and the choice between them is a bet on whether routing succeeds. For incumbents with sunk fleets, the rational strategy is limit pricing above marginal cost to deter entry while refreshing on the power-slot condition --- exactly the behavior Meta Compute and the Colossus leases foreshadow. For policymakers, and for Japan specifically, the luxury--mass segmentation implies that sovereign capability does not require frontier-scale training expenditure: a national position built on efficient inference of open weights, bandwidth-first architectures, and selective premium capability access captures most economic value at a small fraction of the capital cost --- provided the infrastructure is domestically owned and operated, since the CLOUD Act places data on US-controlled infrastructure within US legal reach regardless of siting, and provided AI4SIS workloads (Section 7) are cultivated as the anchor demand that keeps such sovereign capacity utilized independently of enterprise budget cycles --- while the Bifurcation scenario, at 12\%, is precisely the branch in which such sovereign open-weight capacity appreciates most.

The single most decision-relevant observable over the next four quarters is realized token-demand growth net of efficiency --- measurable from public serving-price and volume data --- against the ~1.6--2.4$\times$ threshold band. Secondary indicators: HBM contract-price normalization timing; the absolute-vs-coupled behavior of premium API pricing as routing spreads; and whether announced 2028--2029 capacity FIDs proceed, shrink, or convert to lease structures.

\appendix
\section{Model Parameters}
\begin{table}[t]\centering\footnotesize
\caption{Model parameters used throughout.}
\begin{tabular}{L{3.3cm}L{7.1cm}L{4.1cm}}
\toprule
\textbf{Parameter} & \textbf{Value} & \textbf{Notes} \\
\midrule
Electricity / PUE & \$0.08/kWh / 1.3 & US data-center average \\
Operations & \$0.25 per GPU-hr & staff, network, maintenance allocated \\
System multiplier & 1.5$\times$ GPU price & chassis, networking, DC shell \\
Depreciation & 4-yr straight line & marginal cost thereafter \\
MBU & 0.50 / 0.55 / 0.60--0.62 & Hopper / Blackwell / Rubin \\
Platforms & H100 3.35 TB/s 700W; H200 4.8 TB/s 700W; B200 8 TB/s 1.0kW; GB300 8 TB/s 1.4kW; Rubin 16 TB/s 1.8kW; '29 refresh 20 TB/s 2.2kW &  \\
Prices, base branch & GB300 '26 \$55k; Rubin '27 \$72k, '28 \$58k, '29 \$64k (refresh) & HBM normalizes from 2028 \\
Prices, shortage branch & '27 \$76k, '28 \$78k, '29 \$86k & HBM elevated to 2030 \\
Frontier training & 2.5$\times$/yr FLOP ambition; \$/FLOP $-$25\%/yr base, $-$10\%/yr shortage; \$1.5B 2026 run &  \\
Good-enough training & \$40M 2026, $-$40\%/yr & RL post-training + distillation on open bases \\
Capacity buildout & +45/40/30/15\%/yr net, 2027--30 & announced capex trajectory \\
Efficiency & 30\%/yr bytes-per-token baseline; 15--45\% stress range & KV compression, sparsity, routing \\
Premium pricing & coupled: 7$\times$ mass; sticky: \$0.40/PB absolute & mass price = floor +30\% margin \\
\bottomrule
\end{tabular}\label{tab:5}
\end{table}
{\footnotesize\itshape Sources for market context: TrendForce and Counterpoint DRAM contract-price surveys (Q1--Q3 2026); CNBC and IEEE Spectrum reporting on the memory shortage; Z.ai GLM-5.2 release materials and third-party evaluations; Google Research TurboQuant publication and independent analyses; Bloomberg and SEC-filed reporting on Meta Compute, the xAI--Anthropic Colossus lease, and CoreWeave agreements. All market figures are paraphrased; the quantitative model and all conclusions are original to this analysis.\par}
\section*{Acknowledgments and AI Disclosure}
This paper was prepared with substantial AI assistance: the quantitative models, figures, and successive drafts were developed by the author working with Anthropic's Claude (Fable 5), under the author's direction and with all analytical choices, scenario judgments, and final text the author's responsibility. The author gratefully credits OpenAI's ChatGPT 5.5, which contributed valuable review comments on multiple drafts --- including the measurement critique of the demand trackers, the tempering of the scenario probabilities, the pricing-regime sensitivity framing, and the contract-status classification of the financing web --- many of which are adopted in Sections 8--9 and Tables 2 and 3. Readers should note the disclosure this implies: the two AI systems that assisted in preparing and reviewing this paper are products of two companies whose commercial prospects the paper analyzes; the author has endeavored to keep the analysis evenhanded, and all errors, estimates, and subjective probabilities remain solely the author's. AI-generated content may contain errors; all load-bearing figures have been checked against the primary sources cited in the References, but readers should verify critical numbers independently before relying on them for investment, procurement, or policy decisions. This paper does not constitute investment advice.

\end{document}